# Coaxial tungsten hot plate-based cathode source for Cesium plasma production confined in MPD device

A. D. Patel[1], Zubin Shaikh[2], M. Sharma[1], Santosh P. Pandya[1], and N. Ramasubramanian[1,3]

[1] *Instittute for Plasma Research, Bhat, Gandhinagar, Gujarat-382428, India*

[2] *Department of Physics, Saurastra University, Rajkot, Gujarat-360005, India*

[3] *Homi Bhabha National Institute, Training School Complex, Anushaktinagar, Mumbai 400094, India*

E-mail: amitphy9898@gmail.com

**Abstract:**

A Multi-dipole line cusp configured Plasma Device (MPD) having six electromagnets with embedded Vacoflux-50 as a core material and a hot filament-based cathode for Argon plasma production has been characterized by changing the pole magnetic field values. For the next step ahead, a new tungsten ionizer plasma source for contact ionization cesium plasma has been designed, fabricated, and constructed and thus plasma produced will be confined in MPD. An electron bombardment heating scheme at high voltage is adopted for heating of 6.5cm diameter tungsten plate. This article describes the detailed analysis of the design, fabrication, operation, and characterization of temperature distribution over the tungsten hot plate using the Infrared camera of the tungsten ionizer. The tungsten plate has sufficient temperature for the production of Cesium ions/plasma.

**Design Consideration:**

A tungsten ionizer or a hot plate assembly is an important part of a Q-machine and it is a source of steady plasma production [1]. In this article, the detailed description of the structure and design of an electron bombardment heating mechanism for heating the tungsten material-based plate for contact ionization Cesium plasma or Cesium ion production has been discussed. This coaxial tungsten hot plate-based hot cathode source is capable of producing an electron-emitting surface (6.5cm in diameter) at 2500K. The coaxial structure of the hot cathode allows access along the axis is assembled from modular components and requires no water cooling internal to the vacuum system. The detailed description of the tungsten hot plate-based ionizer is as follows.

- **Material selection**

A tungsten ionizer is a Cesium plasma-producing device operated at high temperatures. The material of different parts must be capable to work at high temperatures and high D. C. voltage. Material selection is a very important procedure as the machine is made to work for longer life at high temperatures. Some crucial factors that should be considered while selecting the material are as follows.

1. Material should withstand against high temperature.
2. All parts are assembled very accurately. A little expansion in any part due to high temperature creates more problems. So, materials should have a very low thermal coefficient of expansion.
3. All materials should be easily workable on the lathe machine.

4. Materials of the parts which are used for heating should be good electric conductors at a higher temperature.
   5. The insulating material has a low heat transfer coefficient.
   6. Materials should resist oxidation and should not react with any gas as plasma is produced through this assembly.
   7. Materials should be easily available.
   8. Cost is also an important consideration while selecting materials.

- **Filament material**

Normally tungsten filaments are used. In some big devices, tantalum filaments have been used to minimize the effect of heating. However, tungsten has been used in smaller devices and the tungsten filaments have a longer lifetime [2].

- **Methods of heating of tungsten ionizer**

A hot plate, the main part of the assembly, is heated up to a high temperature of about 2700 K. Following are some of the methods available for the heating of the hot plate.

   1. In this method, the tungsten plate has been heated by passing a huge current through it. But this method creates troubles in the plasma. A huge current passing through the plate produces the magnetic field and it deforms the ambient magnetic field, as well as the ripple in the power supplies, which may provoke unwanted noise in the plasma.
   2. In this method the plate has been heated by radiation. But it is very difficult to achieve high temperature by radiation.
   3. Another method is to heat the plate by electron bombardment. This electron bombardment is done from the thermionic emission of electrons from the heated tungsten filaments, located behind the plate. Filaments are heated by the passing currents from the floating electric power supply. Filaments may be arranged in parallel connections for the low voltage drop across the filaments [1]. Hence this method is chosen, but it has the following design issue.

- **Problems in design of electron bombardment system**

The problem with making the electron bombardment system for heating the plate is to provide high-temperature and high voltage insulation to the filaments as well as other parts of the assembly. In the assembly, the insulators are placed very nearby the tungsten filaments as well as the plate; a huge temperature will melt the insulators. And if the insulators are placed away from the tungsten filaments, it limits the accuracy as well as alignment of modular parts of the assembly and which is necessary for achieving uniform temperature distribution over the plate [1].

- **Temperature uniformity**

To achieve temperature uniformity over the plate during the operation of tungsten ionizer, following points has been considered:
   1. To blur out the discrete filament structure, the tungsten plate has a thickness (13 mm) that is comparable to or greater than the filament spacing [1].
   2. Space charge limited operation to prevent thermal runaway, as well as precise filament plane alignment with the hot plate. Even if the plate as a whole was stable, local thermal run-away would result in temperature gradients. This problem is avoided by using a space charge limited operation, although it necessitates exceptional alignment precision.

The electron bombardment current density is given by the Child-Langmuir formula,

$$j = \sqrt{\frac{2e}{m}} \frac{\sqrt{V^3}}{9\pi d^2}$$

Where e, m, V and is the electron charge, electron mass, bombardment voltage and filament-plate spacing respectively.

- **3-D Modelling**

A 3-D model is shown in the figure. Figure 1 shows 3-D modelling of tungsten ionizer assembly and becomes crucial to understanding different parts and coaxial arrangement of assembly. Here, modelling is done with the help of modelling software CATIA, available at the institute. A 3-D model is shown in the figure.

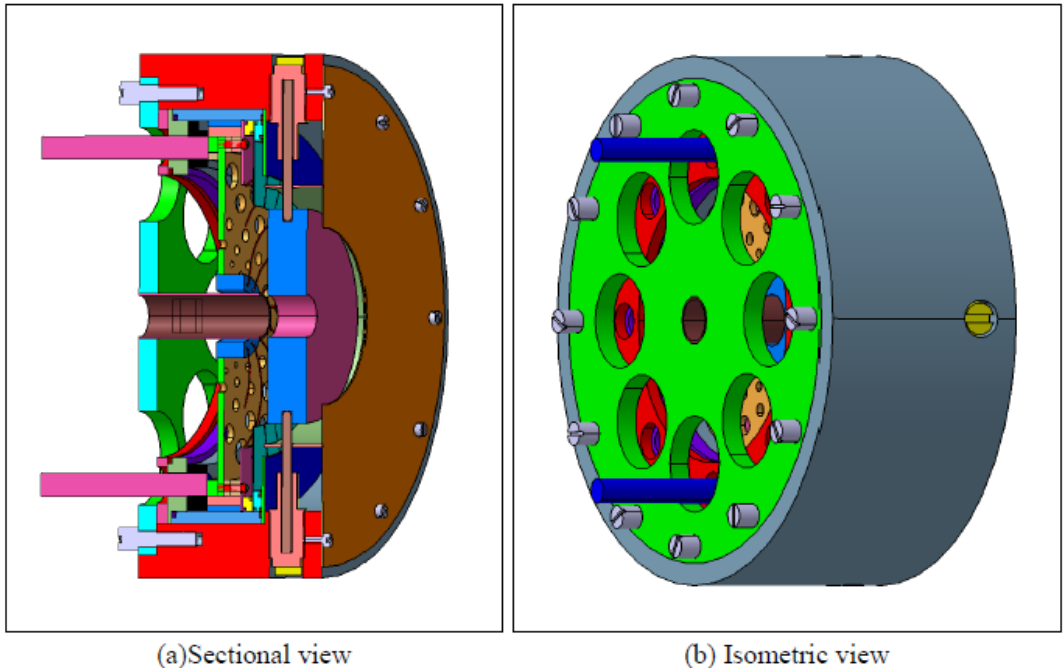

(a) Sectional view        (b) Isometric view
Figure 1: 3-D model of tungsten ionizer in CATIA

- **Assembly**

All parts are assembled one by one as per the assembly procedure. The sectional view A-A of the assembly is shown below.

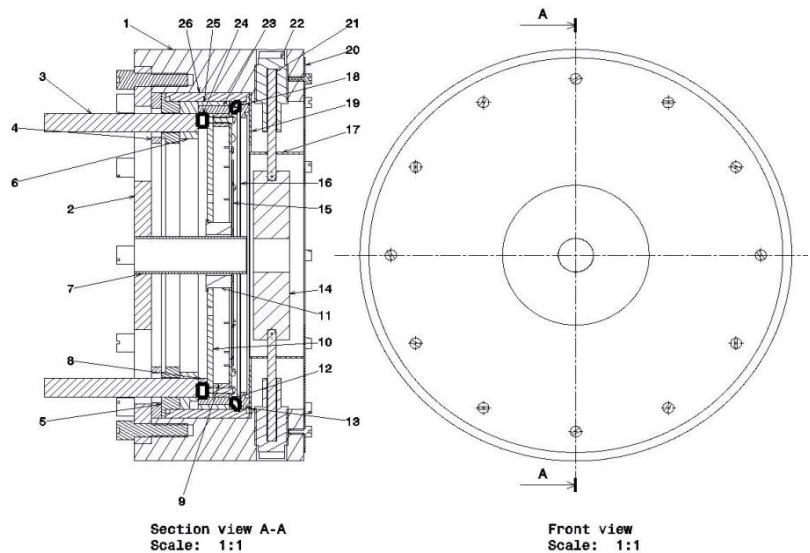

Section view A-A          Front view
Scale: 1:1                Scale: 1:1

In the modelling, all the parts are shown with different colours to make the visualization and identification of parts better and easier. All parts shown in assembly sectional view A-A are listed in the following table 1

| Sr. No. | Part Name | Part No. (As per 2-D sketch) | Material in Prototype | Quantity |
|---|---|---|---|---|
| 1 | Cylinder | 1 | Copper | 1 |
| 2 | Clamping plate | 2 | Copper | 1 |
| 3 | Rod | 3 | Molybdenum | 2 |
| 4 | Spacer | 4 | Copper | 1 |
| 5 | Insulator | 5, 9, 12, 13, 23, 24, 26 | Boron Nitride | 1 |
| 6 | Ring | 6, 8, 25 | Molybdenum | 1 |
| 7 | Tube | 7 | Molybdenum | 1 |
| 8 | Back plate | 10 | Molybdenum | 1 |
| 9 | Inner support ring | 11 | Tantalum | 1 |
| 10 | Outer support ring | 18 | Molybdenum | 1 |
| 11 | Hot plate | 14 | Tungsten | 1 |
| 12 | Filament | 15 | Tungsten | 16 |
| 13 | Heat shield | 17, 19 | Molybdenum | 1 |
| 14 | Floating shield | 16 | Molybdenum | 1 |
| 15 | Aperture limiter | 20 | Molybdenum | 1 |
| 16 | Leg | 21 | Molybdenum | 4 |
| 17 | Socket | 22 | Molybdenum | 4 |
| 18 | M2 bolt | - | Molybdenum | 12 |
| 19 | M3 bolt | - | Molybdenum | 12 |
| 20 | M5 bolt | - | Molybdenum | 12 |

Table 1. Details of parts used in tungsten hot plate based ionizer.

- **Operation & Characterization of hot plate temperature distribution**

The whole tungsten ionizer assembly is placed in a vacuum chamber and fitted at the one end of the chamber and supported by a canonical chamber. The vacuum chamber is pumped out using TMP (turbo-molecular pump) back up by the rotary pump and capable up to $1 \times 10^{-7}$ mbar vacuum. The vacuum chamber outer surface is also actively water-cooled during the operation. Figure 2 shows an electric circuit for the bombardment of electrons on a tungsten plate. After achieving the vacuum $1\times10^{-7}$ mbar, then as per the circuit diagram in figure 2, 300A current pass through the 16 filaments (connected in parallel), step by step up from the 15V, 500A floating power supply (up to 5KV isolation). Figure 3 (a) shows the image of 16 filaments connected in parallel in the assembly and (b) shows the light glow filaments inside the chamber.

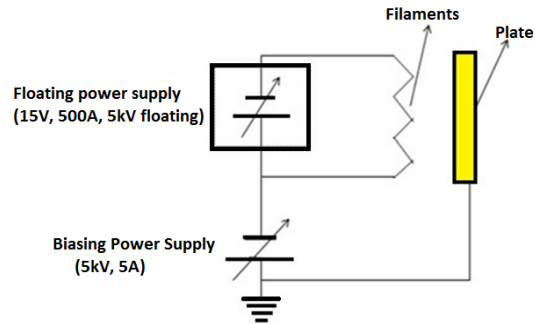

Figure 2: An electric circuit for bombardment of electron on tungsten plate

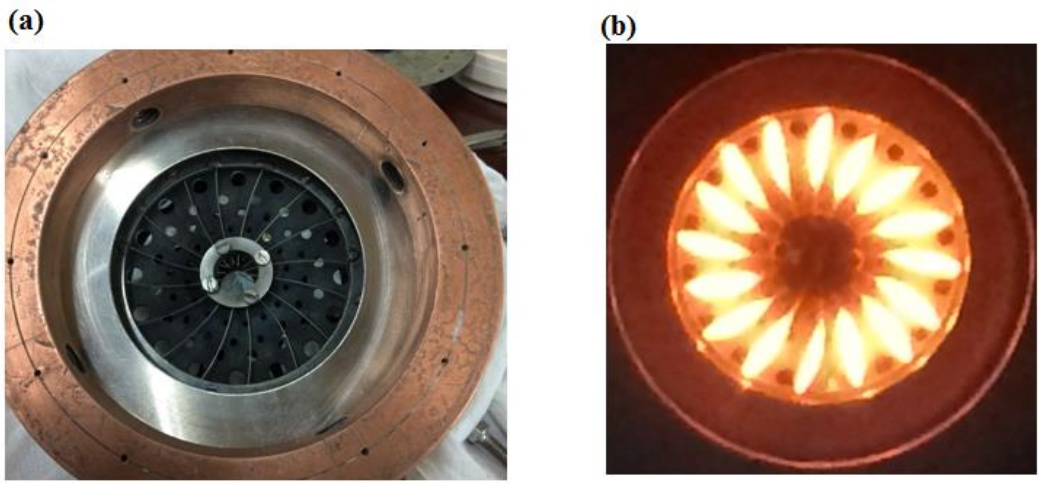

Figure 3: Front view of coaxial cathode source showing filaments arrangements (a) outside the chamber (b) image of coaxial circular plasma source inside the vacuum chamber when 300A current passed through the filaments.

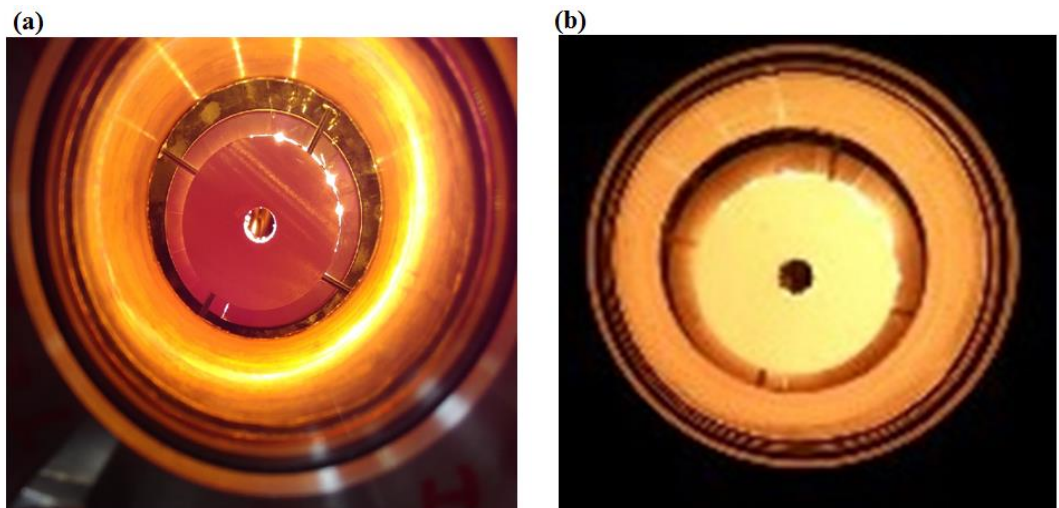

Figure 4: Front view of hot plate (a) hot plate before applying accelerating voltage and (b) after applying 1kV accelerating voltage.

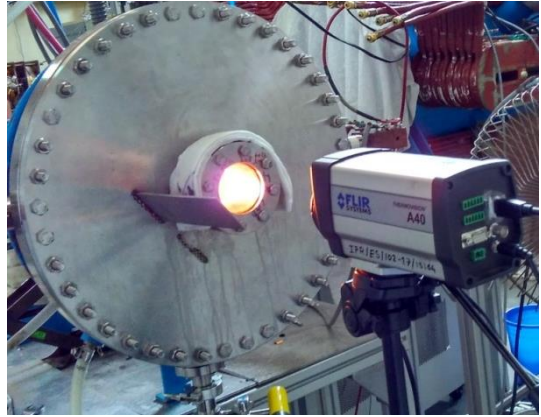

Figure 5: Set up of IR camera in lab to measure temperature distribution over the tungsten hot plate

     Figure 4 (a) shows the front view of the hot plate inside the vacuum chamber when 300A current is passed from the filament and no accelerating voltage is applied and (b) shows the heated plate when 1kV accelerating voltage is applied between plate and filaments. For the temperature distribution analysis over the surface of the hot plate, a non-contact Infrared imaging measurement was used as shown in figure 5. FLIR® make Long Wave Infrared (LWIR) camera [3] (a full-frame rate up to 60 Hz, 320 x 240 pixels array, 8 – 12 μm spectral response range, Field of View 24° x 18°, Temperature sensitivity (NETD) ~0.08 °C, and temperature range 200°C – 1500 °C was used for measurement of temperature distribution of the hot plate. IR-camera is placed outside the vacuum vessel, and observed the hot plate through an infrared transmitting vacuum viewport. Figure 6 (a) shows an IR image of a hot plate showing temperature distribution over the surface of the hot plate when 300 A current is passed through filaments and 1kV accelerating voltage is applied between hotplate and filament for the bombardment of electrons to the plate, and (b) shows the temperature variation of the hot plate along the X-axis. The temperature over the plate is nearly uniform (~1400 K) means the variation in the temperature is nearly ±1% uniform along the X-axis as shown in figure 6(a) as dash line. Because of the limiting range of temperature measurement of IR camera, IR image of higher accelerating voltage was not taken. Figure 7 (b) shows the temperature variation along the Y-axis. It clearly captures a sharp gradient in temperature distribution along the Y direction as shown in figure 7(a) as a dash line. This small gradient in temperature may excite drift wave instability. Besides the tungsten plate has sufficient temperature for the production of Cesium [1].

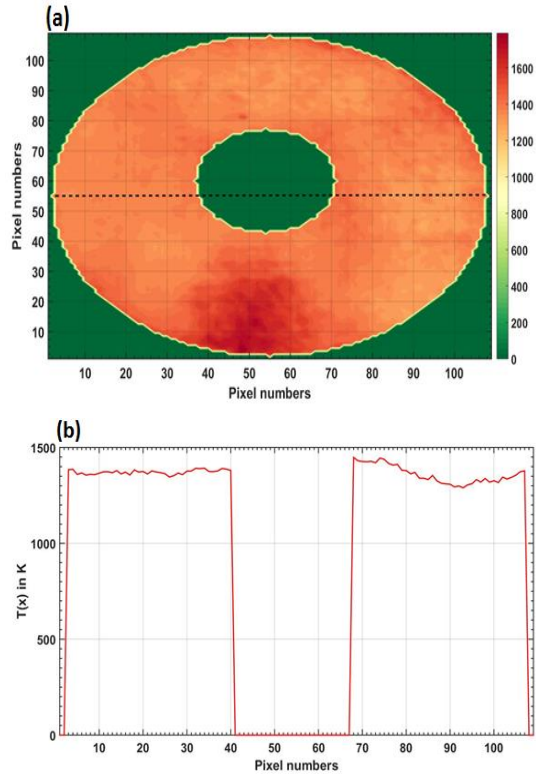

Figure 6 (a) IR image of hot plate and (b) Temperature variation of hot plate surface along X-axis as shown dash line in figure (6(a)), when filament current is 300A and biased accelerating voltage is 1kV.

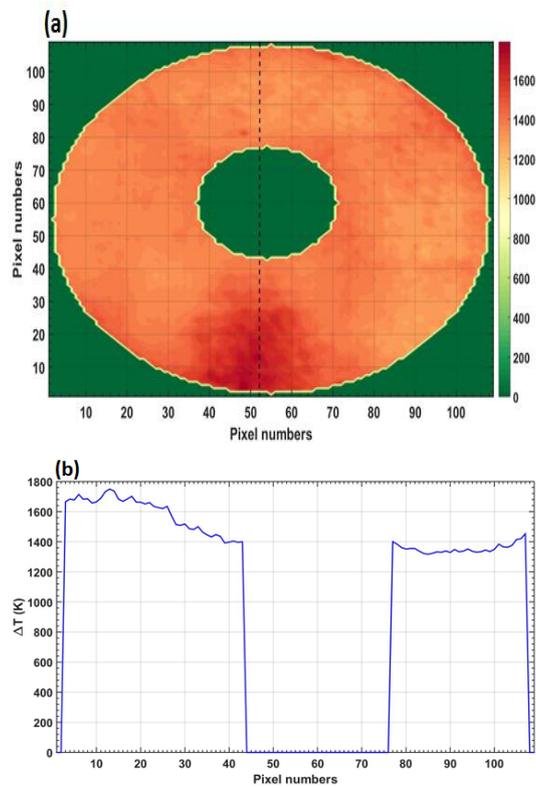

Figure 7 (a) IR image of hot plate (b) Temperature variation of hot plate surface along Y-axis as shown in dash line in figure (7(a)), when filament current is 290A and biased accelerating voltage is 1kV.

- **Conclusion**

Coaxial tungsten hot plate-based cathode for Cesium plasma production has been designed, constructed, and fabricated and the temperature distribution over the plate has been characterized using an IR camera. The following points have been remarkable for this source:

1. The uniform temperature distribution over the tungsten plate determines the uniform plasma density and plasma potential.
2. Strong ambient magnetic fields cause J X B forces and distort the electron-emitting tungsten filaments and thus nearly field-free region of multi-pole cusp magnetic field configuration plays a crucial role in placing the source.
3. A high accelerating voltage causes the arcing and reduced the lifetime of tungsten filaments.
4. IR images clearly capture the temperature gradient over the tungsten plate. Hence new improved design is necessary for better confinement and a better quiescent level of contact ionized alkali plasma.
5. The tungsten plate has sufficient temperature for the production of Cesium ion/plasma.

**Reference:**

1. Francis F. Chen, Coaxial Cathode Design for Plasma Sources, Princeton University, 1969

2. Y. C. Saxena, Double Plasma Device: An Introduction, PRL, Ahmadabad, 1983

3. Infrared Camera Model ThermoVision™A40M, Make FLIR®, LWIR-camera: http://alacron.com/clientuploads/directory/Cameras/FLIR/A40M%20Researcher%20Datasheet.pdf